\documentclass[preprint, superscriptaddress,showpacs,preprintnumbers,amsmath,amssymb]{revtex4}
\usepackage{graphicx}
\usepackage{amsmath}\usepackage{slashed}

\begin{document}

\thispagestyle{empty}

\title{Special features of the thermal Casimir effect across a uniaxial
anisotropic film}

\author{
V.~M.~Mostepanenko}
\affiliation{Central Astronomical Observatory at Pulkovo of the Russian Academy of Sciences,
Saint Petersburg,
196140, Russia}

\affiliation{Institute of Physics, Nanotechnology and
Telecommunications, Peter the Great Saint Petersburg 
Polytechnic University, Saint Petersburg, 195251, Russia}

\begin{abstract}
We investigate the thermal Casimir force between two parallel plates made of
different isotropic materials which are separated by a uniaxial anisotropic film.
Numerical computations of the Casimir pressure at $T=300\,$K are performed using
the complete Lifshitz formula adapted for an anisotropic intervening layer and
in the nonrelativistic limit. It is shown that the standard (nonrelativistic)
theory of the van der Waals force is not applicable in this case, because the
effects of retardation contribute significantly even for film thicknesses of a few
nanometers. We have also obtained simple analytic expressions for the classical
Casimir free energy and pressure for large film thicknesses (high temperatures).
Unlike the case of isotropic intervening films, for two metallic plates the
classical Casimir free energy and pressure are shown to depend on the static
dielectric permittivities of an anisotropic film. One further interesting feature
is that the classical limit is achieved at much shorter separations between the
plates than for a vacuum gap. Possible applications of the obtained results are
discussed.
\end{abstract}

\maketitle

\section{Introduction}
The Casimir force is caused by the zero-point and thermal fluctuations of
the electromagnetic field. It acts between closely spaced material bodies
and becomes dominant at separations below a micrometer. Although the
Casimir force, which is a relativistic analogue of the van der Waals force,
was predicted long ago \cite{1}, it has become the subject of intensive
experimental and theoretical study only recently (see Refs.~\cite{2,3,4,5}
for a review). This was partially stimulated by the promising applications
to nanotechnology.

The most part of research in the Casimir physics was done for isotropic
test bodies separated either with a vacuum gap or a gap filled by an
isotropic material.
Some attention, however, was also devoted to the role
of materials anisotropy. Here, one should mention the pioneer papers
\cite{6,7,8} where the Lifshitz theory of the van der Waals and Casimir
forces \cite{9} was generalized for the case of 3-layer planar systems
made of uniaxial anisotropic materials with one common optical axis,
but only in the nonrelativistic limit.
The formalism of temperature Green's functions was formulated for
anisotropic media in Ref.~\cite{10}. The fully relativistic Lifshitz
formula at nonzero temperature for two parallel plates made of uniaxial
anisotropic materials separated by a uniaxial medium with all three
optical axes perpendicular to the plates was presented in Ref.~\cite{11}.
It was applied to calculate the Hamaker constant in the nonrelativistic
limit. The Casimir torque and the Casimir force for two parallel plates
immersed in liquid or separated by a vacuum gap was considered \cite{12,13}
for parallel (in-plane) and perpendicular to the plates optical axes
(see also Ref.~\cite{14} for the case of in-plane axes).
Note that electromagnetic theory for the plane-parallel layered structure
possessing the most general type of anisotropy was developed in Ref.~\cite{15}.
The Casimir force between both uniaxial and biaxial anisotropic
magnetodielectric materials through a vacuum gap was considered in
Ref.~\cite{16}. The Casimir-Polder force between a polarizable microparticle
and a plate made of the unidirectional crystal of graphite was
calculated in Ref.~\cite{17}. The effect of a nonzero tilt between the
optical axis and the surface normal on the van der Waals force in the
configuration of two parallel plates separated by a vacuum gap was
investigated in Ref.~\cite{18}. All calculations were performed at nonzero
temperature, but in the nonrelativistic limit. Finally, it was shown that
the Casimir force between atomically thin Au films across a vacuum gap
can also be described in the same way as between uniaxial anisotropic
materials \cite{19}.

In this paper, we consider special features of the thermal Casimir force
acting between two parallel plates made of isotropic materials, but
separated with a gap filled by a uniaxial anisotropic dielectric
(BeO, as an example). The optical axis of the latter is assumed to be
perpendicular to the plates. Additional importance of this configuration
is caused by the role it plays in the investigation of stability of
strongly confined liquid crystals \cite{20}. The 3-layer systems are also often
discussed in connection with the Casimir repulsion \cite{4,5,12}.
We perform all calculations at nonzero (room) temperature in the fully
relativistic case. We also obtain the analytic results and perform
computations in the nonrelativistic limit.
One of our main results is that for a gap filled by an
anisotropic material the relativistic effects become essential at much
shorter separations, than it was believed before. In fact, we show that
there is no separation region where the force follows the nonrelativistic
van der Waals regime. This is in contradiction with the statement \cite{18}
that the force between surfaces in uniaxial anisotropic media can be
calculated in the nonretarded limit up to separation distances of
approximately $1\,\mu$m. We also show that the classical limit of the
Casimir interaction is achieved at much shorter separations
than for a vacuum gap. Finally, we obtain simple analytic expressions for the
classical Casimir free energy and pressure and demonstrate that for a gap
filled by an anisotropic material these quantities depend on the
dielectric permittivities of the gap material. This is not the case for
a gap between metallic plates filled by an isotropic substance.

The paper is organized as follows. In Sec.~II we briefly present the
Lifshitz formulas and the reflection coefficients adapted for the
configuration of two isotropic plates interacting across a uniaxial
anisotropic film. We also derive the analytic expressions for the
Casimir free energy and pressure in the nonrelativistic limit.
Section III is devoted to numerical computations  of the Casimir
pressure in the fully relativistic case
and to comparison with respective computational results
in the nonrelativistic limit. Here, we consider three different pairs
of plates (dielectric-dielectric, metallic-metallic and
dielectric-metallic). We derive the classical limit for both
the Casimir free energy and pressure in Sec.~IV and compare the
analytical results with the results of numerical computations.
In Sec.~V the reader will find
our conclusions and discussion.

\section{The Lifshitz formulas for the Casimir free energy and pressure
across a uniaxial anisotropic film}

We consider two thick plates (semispaces) described by the frequency-dependent
dielectric permittivities $\varepsilon^{(-1)}(\omega)$ and
$\varepsilon^{(+1)}(\omega)$ which are separated by the dielectric film of
thickness $a$ at temperature $T$ in thermal equilibrium with an environment.
The material of the film is a uniaxial anisotropic crystal with the optical
axis perpendicular to the plates. We choose the coordinate plane $(x,y)$
parallel to the plates. Then the film material is described by the
frequency-dependent dielectric permittivities
$\varepsilon_{xx}^{(0)}(\omega)=\varepsilon_{yy}^{(0)}(\omega)$ and
$\varepsilon_{zz}^{(0)}(\omega)$.

The Lifshitz formula for the Casimir free energy per unit area in the
configuration of a uniaxial anisotropic film sandwiched between two
isotropic semispaces can be derived, e.g., using the method of surface
modes \cite{5} or the scattering approach \cite{16}. In the fully relativistic
case at $T\neq 0$ this formula is contained in Ref.~\cite{11}.
Using more modern notations typical for the scattering theory it can be
written in a more transparent way as
\begin{eqnarray}
&&
{\cal F}(a,T)=\frac{k_BT}{2\pi}\sum_{l=0}^{\infty}
{\vphantom{\sum}}^{\prime}\int_{0}^{\infty}k_{\bot}\,dk_{\bot}
\label{eq1} \\
&&
\times\left\{\ln\left[1-r_{\rm TM}^{(0,+1)}(i\xi_l,k_{\bot})
r_{\rm TM}^{(0,-1)}(i\xi_l,k_{\bot})
e^{-2ak_{\rm TM}^{(0)}(i\xi_l,k_{\bot})}\right]\right.
\nonumber \\
&&
+\left.\ln\left[1-r_{\rm TE}^{(0,+1)}(i\xi_l,k_{\bot})
r_{\rm TE}^{(0,-1)}(i\xi_l,k_{\bot})
e^{-2ak_{\rm TE}^{(0)}(i\xi_l,k_{\bot})}\right]\right\}.
\nonumber
\end{eqnarray}
\noindent
Here, $k_B$ is the Boltzmann constant,
$k_{\bot}=|\mbox{\boldmath$k$}_{\bot}|$ is the magnitude of the
projection of the wave vector on the plane of plates, the prime on
the summation sign multiples the term with $l=0$ by 1/2,
$\xi_l=2\pi k_BTl/\hbar$ with $l=0,\,1,\,2,\,\ldots$ are the
Matsubara frequencies, and the quantities $k^{(0)}$ for two independent
polarizations of the electromagnetic field, transverse magnetic (TM) and
transverse electric (TE), are given by
\begin{eqnarray}
&&
k_{\rm TM}^{(0)}(i\xi_l,k_{\bot})=\sqrt{
\frac{\varepsilon_{xx,l}^{(0)}}{\varepsilon_{zz,l}^{(0)}}k_{\bot}^2+
{\varepsilon_{xx,l}^{(0)}}\frac{\xi_l^2}{c^2}},
\nonumber \\
&&
k_{\rm TE}^{(0)}(i\xi_l,k_{\bot})=\sqrt{
k_{\bot}^2+
{\varepsilon_{xx,l}^{(0)}}\frac{\xi_l^2}{c^2}},
\label{eq2}
\end{eqnarray}
\noindent
where $\varepsilon_{xx,l}^{(0)}\equiv\varepsilon_{xx}^{(0)}(i\xi_l)$ and
$\varepsilon_{zz,l}^{(0)}\equiv\varepsilon_{zz}^{(0)}(i\xi_l)$.
Note that in the case of a uniaxial crystal with the optical axis
perpendicular to the plates the two polarizations of the electromagnetic
field separate. This is, however, not so for media with the most general
type of anisotropy \cite{15}.

The quantities $r_{\rm TM,TE}^{(0,\pm 1)}$ in Eq.~(\ref{eq1}) are the
reflection coefficients for the TM and TE polarizations.
The reflection coefficients on the interface of a uniaxial and isotropic
media were derived long ago \cite{21} (see also Refs.~\cite{5,16,17,22}).
They are given by
\begin{eqnarray}
&&
r_{\rm TM}^{(0,\pm 1)}(i\xi_l,k_{\bot})=\frac{\varepsilon_{l}^{(\pm 1)}
k_{\rm TM}^{(0)}(i\xi_l,k_{\bot})-\varepsilon_{xx,l}^{(0)}
k^{(\pm 1)}(i\xi_l,k_{\bot})}{\varepsilon_{l}^{(\pm 1)}
k_{\rm TM}^{(0)}(i\xi_l,k_{\bot})+\varepsilon_{xx,l}^{(0)}
k^{(\pm 1)}(i\xi_l,k_{\bot})},
\nonumber \\
&&
r_{\rm TE}^{(0,\pm 1)}(i\xi_l,k_{\bot})=\frac{k_{\rm TE}^{(0)}(i\xi_l,k_{\bot})-
k^{(\pm 1)}(i\xi_l,k_{\bot})}{k_{\rm TE}^{(0)}(i\xi_l,k_{\bot})+
k^{(\pm 1)}(i\xi_l,k_{\bot})},
\label{eq3}
\end{eqnarray}
\noindent
where $\varepsilon_{l}^{(\pm 1)}\equiv\varepsilon^{(\pm 1)}(i\xi_l)$ and
\begin{equation}
k^{(\pm 1)}(i\xi_l,k_{\bot})=\sqrt{
k_{\bot}^2+
{\varepsilon_{l}^{(\pm 1)}}\frac{\xi_l^2}{c^2}}.
\label{eq4}
\end{equation}
\noindent
Important characteristic feature of the Lifshitz formula (\ref{eq1}) is that
the quantities $k_{\rm TM}^{(0)}$ and $k_{\rm TE}^{(0)}$ defined in Eq.~(\ref{eq2})
are determined by dissimilar dielectric permittivities and coincide only in the
isotropic limit
\begin{equation}
k_{\rm TM}^{(0)}(i\xi_l,k_{\bot})=k_{\rm TE}^{(0)}(i\xi_l,k_{\bot})=\sqrt{
k_{\bot}^2+
{\varepsilon_{l}^{(0)}}\frac{\xi_l^2}{c^2}}.
\label{eq6}
\end{equation}
\noindent
This make the case of anisotropic intervening material more  rich in physical
consequences, as compared to the case of isotropic one.

The Casimir pressure between two thick isotropic plates separated by the uniaxial
anisotropic film is obtained from Eq.~(\ref{eq1}) by the negative differentiation
with respect to separation
\begin{eqnarray}
&&
P(a,T)=-\frac{\partial{\cal F}(a,T)}{\partial a}=-\frac{k_BT}{\pi}\sum_{l=0}^{\infty}
{\vphantom{\sum}}^{\prime}\int_{0}^{\infty}k_{\bot}\,dk_{\bot}
\label{eq7} \\
&&
\times\left\{k_{\rm TM}^{(0)}(i\xi_l,k_{\bot})\left[
\frac{e^{2ak_{\rm TM}^{(0)}(i\xi_l,k_{\bot})}}{r_{\rm TM}^{(0,+1)}(i\xi_l,k_{\bot})
r_{\rm TM}^{(0,-1)}(i\xi_l,k_{\bot})}-1
\right]^{-1}\right.
\nonumber \\
&&
+\left.
k_{\rm TE}^{(0)}(i\xi_l,k_{\bot})\left[
\frac{e^{2ak_{\rm TE}^{(0)}(i\xi_l,k_{\bot})}}{r_{\rm TE}^{(0,+1)}(i\xi_l,k_{\bot})
r_{\rm TE}^{(0,-1)}(i\xi_l,k_{\bot})}-1
\right]^{-1}
\right\}.
\nonumber
\end{eqnarray}
\noindent
This equation up to different notations coincides with respective result of
Ref.~\cite{11}.

For the purpose of numerical computations, it is convenient to rewrite Eq.~(\ref{eq7})
in terms of dimensionless variables. We introduce the dimensionless Matsubara
frequencies
\begin{equation}
\zeta_l\equiv\frac{\xi_l}{\omega_c}=\frac{2a\xi_l}{c}\equiv\tau l,
\quad
\tau\equiv\frac{4\pi ak_BT}{c\hbar}.
\label{eq8}
\end{equation}
\noindent
We also introduce different dimensionless wave vector variables in the TM and TE
contributions to Eq.~(\ref{eq7}). In the TM contribution we put
\begin{equation}
y=2a\sqrt{\frac{\varepsilon_{zz,l}^{(0)}}{\varepsilon_{xx,l}^{(0)}}}
k_{\rm TM}^{(0)}(i\xi_l,k_{\bot})=2a\sqrt{
k_{\bot}^2+
{\varepsilon_{zz,l}^{(0)}}\frac{\xi_l^2}{c^2}}.
\label{eq9}
\end{equation}
\noindent
This leads to
\begin{equation}
y\geq\sqrt{\varepsilon_{zz,l}^{(0)}}\zeta_l=
\sqrt{\varepsilon_{zz,l}^{(0)}}\tau l.
\label{eq10}
\end{equation}
\noindent
 In the TE contribution to Eq.~(\ref{eq7}) we put
\begin{equation}
y=2a
k_{\rm TE}^{(0)}(i\xi_l,k_{\bot})=2a\sqrt{
k_{\bot}^2+
{\varepsilon_{xx,l}^{(0)}}\frac{\xi_l^2}{c^2}},
\label{eq11}
\end{equation}
\noindent
which results in
\begin{equation}
y\geq\sqrt{\varepsilon_{xx,l}^{(0)}}\zeta_l=
\sqrt{\varepsilon_{xx,l}^{(0)}}\tau l.
\label{eq12}
\end{equation}

As a result, in terms of dimensionless variables Eq.~(\ref{eq7}) takes the form
\begin{eqnarray}
&&
P(a,T)=-\frac{k_BT}{8\pi a^3}\sum_{l=0}^{\infty}
{\vphantom{\sum}}^{\prime}\left\{
\vphantom{\int}
\sqrt{\frac{\varepsilon_{xx,l}^{(0)}}{\varepsilon_{zz,l}^{(0)}}}
\right.
\label{eq13} \\
&&
\times
\int_{\sqrt{\varepsilon_{zz,l}^{(0)}}\zeta_l}^{\infty}y^2dy
\left[
\frac{e^{\sqrt{\varepsilon_{xx,l}^{(0)}}y/\sqrt{\varepsilon_{zz,l}^{(0)}}}}{r_{\rm TM}^{(0,+1)}(i\zeta_l,y)
r_{\rm TM}^{(0,-1)}(i\zeta_l,y)}-1
\right]^{-1}
\nonumber \\
&&
+\left.
\int_{\sqrt{\varepsilon_{xx,l}^{(0)}}\zeta_l}^{\infty}y^2dy
\left[
\frac{e^{y}}{r_{\rm TE}^{(0,+1)}(i\zeta_l,y)
r_{\rm TE}^{(0,-1)}(i\zeta_l,y)}-1
\right]^{-1}
\right\}.
\nonumber
\end{eqnarray}
\noindent
Here, the reflection coefficients depending on the dimensionless variables
are obtained by using Eqs.~(\ref{eq3}), (\ref{eq4}), (\ref{eq8}),
(\ref{eq9}), and (\ref{eq11})
\begin{eqnarray}
&&
r_{\rm TM}^{(0,\pm 1)}(i\zeta_l,y)=\frac{\varepsilon_{l}^{(\pm 1)}y-
\sqrt{\varepsilon_{xx,l}^{(0)}\varepsilon_{zz,l}^{(0)}}
\sqrt{y^2+[\varepsilon_{l}^{(\pm 1)}-\varepsilon_{zz,l}^{(0)}]
\zeta_l^2}}{\varepsilon_{l}^{(\pm 1)}y+
\sqrt{\varepsilon_{xx,l}^{(0)}\varepsilon_{zz,l}^{(0)}}
\sqrt{y^2+[\varepsilon_{l}^{(\pm 1)}-\varepsilon_{zz,l}^{(0)}]
\zeta_l^2}},
\nonumber \\
&&
r_{\rm TE}^{(0,\pm 1)}(i\zeta_l,y)=\frac{y-
\sqrt{y^2+[\varepsilon_{l}^{(\pm 1)}-\varepsilon_{xx,l}^{(0)}]
\zeta_l^2}}{y+
\sqrt{y^2+[\varepsilon_{l}^{(\pm 1)}-\varepsilon_{xx,l}^{(0)}]
\zeta_l^2}}.
\label{eq14}
\end{eqnarray}

For comparison purposes, it is useful also to consider the Casimir free energy
and pressure in the nonrelativistic limit. In this case only the TM contributions to
Eqs.~(\ref{eq1}) and (\ref{eq7}) survive and we find
\begin{eqnarray}
&&
{\cal F}_{\rm nr}(a,T)=\frac{k_BT}{2\pi}\sum_{l=0}^{\infty}
{\vphantom{\sum}}^{\prime}\int_{0}^{\infty}k_{\bot}\,dk_{\bot}
\ln\left[1-r_{{\rm nr},l}^{(0,+1)}r_{{\rm nr},l}^{(0,-1)}
e^{-2ak_{\bot}\sqrt{\varepsilon_{xx,l}^{(0)}}/\sqrt{\varepsilon_{zz,l}^{(0)}}}\right],
\nonumber \\[-1mm]
&&
\label{eq15} \\
&&
P_{\rm nr}(a,T)=-\frac{k_BT}{\pi}\sum_{l=0}^{\infty}
{\vphantom{\sum}}^{\prime}
\sqrt{\frac{\varepsilon_{xx,l}^{(0)}}{\varepsilon_{zz,l}^{(0)}}}
\int_{0}^{\infty}k_{\bot}^2\,dk_{\bot}
\left[
\frac{e^{2ak_{\bot}\sqrt{\varepsilon_{xx,l}^{(0)}}/
\sqrt{\varepsilon_{zz,l}^{(0)}}}}{r_{{\rm nr},l}^{(0,+1)}r_{{\rm nr},l}^{(0,-1)}}
-1\right]^{-1},
\nonumber
\end{eqnarray}
\noindent
where the reflection coefficients (\ref{eq3}) are reduced to
\begin{equation}
r_{{\rm nr},l}^{(0,\pm 1)}\equiv
r_{{\rm nr}}^{(0,\pm 1)}(i\xi_l)=\frac{\varepsilon_l^{(\pm 1)}-
\sqrt{\varepsilon_{xx,l}^{(0)}\varepsilon_{zz,l}^{(0)}}}{\varepsilon_l^{(\pm 1)}
+\sqrt{\varepsilon_{xx,l}^{(0)}\varepsilon_{zz,l}^{(0)}}}.
\label{eq16}
\end{equation}

It is convenient to introduce the new variable
\begin{equation}
y=2a
\sqrt{\frac{\varepsilon_{xx,l}^{(0)}}{\varepsilon_{zz,l}^{(0)}}} k_{\bot}
\label{eq17}
\end{equation}
\noindent in Eq.~(\ref{eq15}). Then the latter can be rewritten in the form
\begin{eqnarray}
&&
{\cal F}_{\rm nr}(a,T)=\frac{k_BT}{8\pi a^2}\sum_{l=0}^{\infty}
{\vphantom{\sum}}^{\prime}
\frac{\varepsilon_{zz,l}^{(0)}}{\varepsilon_{xx,l}^{(0)}}
\int_{0}^{\infty}ydy
\ln\left[1-r_{{\rm nr},l}^{(0,+1)}r_{{\rm nr},l}^{(0,-1)}
e^{-y}\right],
\nonumber \\[-1mm]
&&
\label{eq18} \\
&&
P_{\rm nr}(a,T)=-\frac{k_BT}{8\pi a^3}\sum_{l=0}^{\infty}
{\vphantom{\sum}}^{\prime}
{\frac{\varepsilon_{zz,l}^{(0)}}{\varepsilon_{xx,l}^{(0)}}}
\int_{0}^{\infty}y^2dy
\left[
\frac{e^{y}}{r_{{\rm nr},l}^{(0,+1)}r_{{\rm nr},l}^{(0,-1)}}
-1\right]^{-1}.
\nonumber
\end{eqnarray}
\noindent
The integral entering the free energy can be represented as
\begin{equation}
-\int_{0}^{\infty}ydy\sum_{k=1}^{\infty}\frac{1}{k}
[r_{{\rm nr},l}^{(0,+1)}r_{{\rm nr},l}^{(0,-1)}]^k e^{-ky}
=-\sum_{k=1}^{\infty}\frac{[r_{{\rm nr},l}^{(0,+1)}
r_{{\rm nr},l}^{(0,-1)}]^k}{k^3}=
-{\rm Li}_3[r_{{\rm nr},l}^{(0,+1)}r_{{\rm nr},l}^{(0,-1)}],
\label{eq19}
\end{equation}
\noindent
where ${\rm Li}_{\nu}(z)$ is the polylogarithm function.
In a similar way, the integral entering Eq.~(\ref{eq18}) for the
Casimir pressure is calculated as
\begin{equation}
\int_{0}^{\infty}y^2dy\sum_{k=1}^{\infty}\frac{1}{k}
[r_{{\rm nr},l}^{(0,+1)}r_{{\rm nr},l}^{(0,-1)}]^k e^{-ky}
=2\sum_{k=1}^{\infty}\frac{[r_{{\rm nr},l}^{(0,+1)}
r_{{\rm nr},l}^{(0,-1)}]^k}{k^3}=
2{\rm Li}_3[r_{{\rm nr},l}^{(0,+1)}r_{{\rm nr},l}^{(0,-1)}].
\label{eq20}
\end{equation}
\noindent
Substituting Eqs.~(\ref{eq19}) and (\ref{eq20}) in Eq.~(\ref{eq18}),
we arrive to the analytic expressions for the Casimir free energy and
pressure in the nonrelativistic limit
\begin{eqnarray}
&&
{\cal F}_{\rm nr}(a,T)=-\frac{k_BT}{8\pi a^2}\sum_{l=0}^{\infty}
{\vphantom{\sum}}^{\prime}
\frac{\varepsilon_{zz,l}^{(0)}}{\varepsilon_{xx,l}^{(0)}}
{\rm Li}_3[r_{{\rm nr},l}^{(0,+1)}r_{{\rm nr},l}^{(0,-1)}],
\nonumber \\[-1mm]
&&
\label{eq21} \\
&&
P_{\rm nr}(a,T)=-\frac{k_BT}{4\pi a^3}\sum_{l=0}^{\infty}
{\vphantom{\sum}}^{\prime}
{\frac{\varepsilon_{zz,l}^{(0)}}{\varepsilon_{xx,l}^{(0)}}}
{\rm Li}_3[r_{{\rm nr},l}^{(0,+1)}r_{{\rm nr},l}^{(0,-1)}],
\nonumber
\end{eqnarray}
\noindent
where the reflection coefficients are defined in Eq.~(\ref{eq16}).

\section{Computation of the Casimir pressure and role of the relativistic effects}

Now we compute the Casimir pressure between different pairs of isotropic
plates (dielectric-dielectric, metallic-metallic and dielectric-metallic)
across a uniaxial dielectric film at room temperature $T=300\,$K.
Computations are performed in a wide region of film thicknesses from 1\,nm
(smaller thicknesses are outside the application region of the Lifshitz
theory) to $5\,\mu$m, where, as we show below, the classical regime has
already long been achieved. Preference is given to the Casimir pressure as
to an immediately measured quantity. If the uniaxial crystal film fills the gap,
application of the proximity force approximation \cite{5}, which is often
used to transform the free energy per unit area of parallel plates into
the Casimir force between a sphere and a plate, seems unjustified because
the optical axis of a film material is generally not perpendicular to the
sphere surface.

To perform computations of the Casimir pressure, one needs the dielectric
permittivities of both plates and a film calculated at the imaginary
Matsubara frequencies. As a material of the uniaxial anisotropic film,
we choose BeO whose dielectric response is well described analytically
in the following form \cite{23}:
\begin{equation}
\varepsilon_{xx(zz),l}^{(0)}=1+
\frac{C_{xx(zz)}^{\rm IR}
\left(\omega_{xx(zz)}^{\rm IR}\right)^2}{\xi_l^2+
\left(\omega_{xx(zz)}^{\rm IR}\right)^2}+
\frac{C_{xx(zz)}^{\rm UV}
\left(\omega_{xx(zz)}^{\rm UV}\right)^2}{\xi_l^2+
\left(\omega_{xx(zz)}^{\rm UV}\right)^2}.
\label{eq22}
\end{equation}
\noindent
Here, $C_{xx(zz)}^{\rm IR}$ and $C_{xx(zz)}^{\rm UV}$ are the absorption strengths in
the infrared (IR) and ultraviolet (UV) range for the components $xx$ and $zz$,
respectively. The respective characteristic absorption frequencies are
$\omega_{xx(zz)}^{\rm IR}$ and $\omega_{xx(zz)}^{\rm UV}$.
The values all the above parameters are \cite{23}
$C_{xx}^{\rm IR}=4.04$, $C_{xx}^{\rm UV}=1.90$,
$\omega_{xx}^{\rm IR}=1.3\times 10^{14}\,$rad/s,
$\omega_{xx}^{\rm UV}=1.98\times 10^{16}\,$rad/s;
$C_{zz}^{\rm IR}=4.70$, $C_{zz}^{\rm UV}=1.951$,
$\omega_{zz}^{\rm IR}=1.4\times 10^{14}\,$rad/s,
$\omega_{zz}^{\rm UV}=2.37\times 10^{16}\,$rad/s.
{}From Eq.~(\ref{eq22}) for the static dielectric permittivities we find
$\varepsilon_{xx}^{(0)}(0)=6.94$ and $\varepsilon_{zz}^{(0)}(0)=7.95$.

First we consider the parallel plates made of amorphous SiO${}_2$ (silica).
The dielectric permittivity of silica along the imaginary frequency axis is
also well described analytically \cite{23}
\begin{equation}
\varepsilon_l^{(\pm 1)}=1+\sum_{j=1}^{3}
\frac{C_{j}^{\rm IR}
(\omega_{j}^{\rm IR})^2}{\xi_l^2+
(\omega_{j}^{\rm IR})^2}+
\frac{C^{\rm UV}
(\omega^{\rm UV})^2}{\xi_l^2+
(\omega^{\rm UV})^2},
\label{eq23}
\end{equation}
\noindent
with the following values of all parameters:
$C_{1}^{\rm IR}=0.829$, $C_{2}^{\rm IR}=0.095$, $C_{3}^{\rm IR}=0.798$,
$\omega_{1}^{\rm IR}=0.867\times 10^{14}\,$rad/s,
$\omega_{2}^{\rm IR}=1.508\times 10^{14}\,$rad/s,
$\omega_{3}^{\rm IR}=2.026\times 10^{14}\,$rad/s,
$C^{\rm UV}=1.098$,
$\omega^{\rm UV}=2.034\times 10^{16}\,$rad/s.
Then Eq.~(\ref{eq23}) leads to the static dielectric permittivity of
SiO${}_2$ $\varepsilon^{(\pm 1)}(0)=3.82$.

We substitute Eqs.~({\ref{eq23}) and (\ref{eq24}) in Eq.~{\ref{eq14}) for
the reflection coefficients and calculate the Casimir pressure across a
uniaxial anisotropic film at $T=300\,$K by Eq.~({\ref{eq13}) as a function
of the film thickness. The computational results for the magnitude of the
Casimir pressure multiplied by the third power of film thickness are
presented in Fig.~\ref{fg1} by the solid line 1 for the range of film
thicknesses from 1 to 10\,nm. In the same figure the dashed line 1 shows
the computational results obtained using Eq.~({\ref{eq21}) for
$P_{\rm nr}(a,T)$, i.e., in the nonrelativistic limit.
As can be seen in Fig.~\ref{fg1}, for a gap filled by the anisotropic
material, the nonrelativistic results differ from the exact ones
significantly even for very thin films.
The relative error of the nonrelativistic Casimir pressure (\ref{eq21})
\begin{equation}
\delta P_{\rm nr}=\frac{|P_{\rm nr}|-|P|}{|P|}
\label{eq24}
\end{equation}
\noindent
is equal to 2.6\%, 8.0\%, 30.9\%. and 79.8\% for film thicknesses of
1, 2, 5, and 10\,nm, respectively. For comparison, the same relative error
for two SiO${}_2$ plates interacting across a vacuum gap is equal to
0.98\%, 3.0\%, 12.0\%, and 30.5\% at the same respective separations.
Note that even for a vacuum gap, the application region of the nonrelativistic
Lifshitz formula (the van der Waals regime) is restricted to very short
separations up to several nanometers. As to the gap, filled by an anisotropic
material, the error of the nonrelativistic Casimir pressure becomes too large
even for the smallest film thicknesses, and for $a=10\,$nm the nonrelativistic
result is already rudely incorrect. Thus, our calculations do not support the
statement \cite{18} that the nonretarded limit can be used to calculate the
Casimir force between surfaces in uniaxial anisotropic media up to
separation distance of $1\,\mu$m.

In Fig.~\ref{fg2}(a) we present our computational results over the wide range
of film thicknesses from 10\,nm to $3\,\mu$m (the solid line).
Taking into account that the attractive (negative) Casimir force for large $a$
depends on separation differently than at short separations, here we plot the
Casimir pressure multiplied by the fourth power of separation.
As can be seen in Fig.~\ref{fg2}(a), for film thicknesses larger than approximately
$2\,\mu$m the solid line becomes straight, i.e., coincides with the dashed line
which describes the large separation (high temperature) classical limit.
In this limiting case the Casimir pressure is inverse proportional to $a^3$
(see Sec.~IV for analytic expressions of the Casimir free energy and pressure
across a uniaxial anisotropic film in the classical limit and related discussion).
Now we only note that for the vacuum gap between two plates the classical
limit is achieved at larger separations of about $6\,\mu$m \cite{5}.

Next, we consider two parallel plates made of Au interacting through the same
uniaxial anisotropic film made of BeO. The dielectric permittivity of Au along
the imaginary frequency axis is widely discussed in the literature in
connection with experiments on measuring the Casimir force (see Refs.~\cite{2,5}
for a review). The imaginary part of the dielectric permittivity along the real
frequency axis is usually obtained by using the tabulated optical data for the
complex index of refraction of Au \cite{24} extrapolated down to zero frequency
either by the Drude model or by the plasma model. Then, the values of
$\varepsilon^{(+1)}=\varepsilon^{(-1)}$ at the imaginary Matsubara frequencies
are found by means of the Kramers-Kronig relation. Note that extrapolations
using the plasma frequency of Au $\omega_p=9.0\,$eV were found to be in excellent
agreement with the dielectric permittivities of Au at the imaginary Matsubara
frequencies found by means of the weighted Kramers-Kronig relations based solely
on the measured optical data \cite{25}.
{}From the standard theoretical point of view, the use of the Drude model
extrapolation to low frequencies is considered as preferable.
However, the experimental data of many precise experiments on measuring the
Casimir interaction between metallic surfaces performed by different experimental
groups are consistent with the plasma model extrapolation and exclude the Drude
model extrapolation \cite{2,5,26,27,28,29,30,31,32}. The measure of consistency
achieves 90\% \cite{33} and the measure of exclusion is as high as up to 99.9\%
\cite{28}. Some doubts were remaining until very recently because in all these
experiments, performed at separations below a micrometer, the difference in
theoretical predictions using different extrapolations does not exceed a few
percent of the measured quantity.
Important breakthrough was achieved after the publication of Ref.~\cite{34}
(see also Ref.~\cite{35}), which proposed the differential experiment where
theoretical predictions using the Drude and the plasma model extrapolations differ
by a factor of 1000. The recently reported first data sets of this experiment
are in agreement with the plasma model extrapolation and exclude the Drude model
one \cite{36}.

Taking into account the above discussion, we use in our computations the
dielectric permittivity of Au obtained by using the tabulated optical data \cite{24}
extrapolated to low frequencies by means of the plasma model \cite{5}.
The magnitudes of the Casimir pressure multiplied by $a^3$ are computed using
Eqs.~(\ref{eq13}) and (\ref{eq14}) and are shown by the solid line 2 in Fig.~\ref{fg1}.
The dashed line 2 presents the nonrelativistic results for the Casimir pressure
between two Au plates across a gap filled by the uniaxial anisotropic material.
These results are obtained by using Eqs.~(\ref{eq15}) and (\ref{eq16}).
As is seen in Fig.~1, the nonrelativistic results, i.e., the nonretarded
van der Waals force, deviate significantly from the exact results even at the
shortest separations between the plates. Thus, at $a=1$, 2, 5, and 10\,nm
the relative error (\ref{eq24}) in the nonrelativistic pressure is equal to
2\%, 4.9\%, 13.9\%, and 27.6\%. Although these errors are smaller than
those in the
case of dielectric plates, they demonstrate that even for very thin films
the relativistic effects contribute essentially.

In Fig.~\ref{fg2}(b) we plot the Casimir pressure between two Au plates
across a uniaxial BeO film multiplied by the fourth power of separation over
the wide separation region from 10\,nm to $3\,\mu$m (the solid line).
As is seen from Fig.~\ref{fg2}(b), in the region from 500 to 1000\,nm
the solid line demonstrates nearly constant behavior, i.e., the characteristic
for metallic plates dependence of the Casimir pressure $\sim a^{-4}$.
At separations above approximately $2.5\,\mu$m the solid line merges with the
straight dashed line demonstrating the characteristic behavior of the Casimir
pressure at large separations (high temperatures), i.e., the classical limit
(see the analytic results for the Casimir free energy and pressure in this
case in Sec.~IV). Similar to the case of dielectric plates, here the classical
limit is achieved at much shorter separations between the plates than for a
vacuum gap.

Finally, we briefly discuss the case of dissimilar plates, i.e., one plate made
of dielectric (SiO${}_2$) and another one made of metal (Au).
It is common knowledge that the 3-layer systems with some relationships between
the dielectric permittivities of the layers demonstrate the effect of the Casimir
repulsion \cite{4,5,12,13}, which was observed experimentally \cite{37} in the
case of an isotropic liquid film sandwiched between two isotropic plates.
In our case of two dissimilar plates and a gap filled by an anisotropic
material the static dielectric permittivities satisfy the inequalities
\begin{equation}
\varepsilon^{(-1)}(0)<\varepsilon_{xx}^{(0)}(0)<
\varepsilon_{zz}^{(0)}(0)<\varepsilon^{(+1)}(0)=\infty
\label{eq25}
\end{equation}
\noindent
and one should expect the effect of repulsion. This expectation is confirmed by
the computational results.

In Fig.~\ref{fg3} we present on the logarithmic scale the values of the Casimir
pressure computed by Eqs.~(\ref{eq13}) and (\ref{eq14}) for the film thicknesses
from 1\,nm to $5\,\mu$m. As is seen from Fig.~\ref{fg3}, the Casimir pressure is
positive over the entire range of film thicknesses which corresponds to the Casimir
repulsion. At separations above approximately $2\,\mu$m the Casimir pressure
between two dissimilar plates across a uniaxial anisotropic gap becomes
classical, i.e., behaves as $\sim a^{-3}$. This region is shown by the dashed
line in Fig.~\ref{fg3}. It is discussed in more details in the next section.

\section{The classical Casimir effect across a uniaxial anisotropic film}

Here, we derive simple analytic expressions for the Casimir free energy and
pressure between two isotropic plates separated by a uniaxial dielectric film
in the case of large film thicknesses (high temperatures). In this case all
terms with $l\geq 1$ in Eqs.~(\ref{eq1}) and (\ref{eq7}) are exponentially
small and only the terms with $l=0$ determine the total result \cite{5}.
{}From Eqs.~(\ref{eq2})--(\ref{eq4}) in the case of two dielectric plates
we obtain
\begin{eqnarray}
&&
r_{\rm TM}^{(0,\pm 1)}(0)=\frac{\varepsilon_{0}^{(\pm 1)}
-\sqrt{\varepsilon_{xx,0}^{(0)}\varepsilon_{zz,0}^{(0)}}}{\varepsilon_{0}^{(\pm 1)}
+\sqrt{\varepsilon_{xx,0}^{(0)}\varepsilon_{zz,0}^{(0)}}}
\nonumber \\
&&
r_{\rm TE}^{(0,\pm 1)}(0)=0.
\label{eq26}
\end{eqnarray}

Substituting Eq.~(\ref{eq26}) in Eqs.~(\ref{eq1}) and (\ref{eq7}) restricted to
the term $l=0$, one obtains
\begin{eqnarray}
&&
{\cal F}_{\rm cl}(a,T)=\frac{k_BT}{4\pi}\int_{0}^{\infty} k_{\bot}dk_{\bot}
\ln\left[1-r_{\rm TM}^{(0,+1)}(0)r_{\rm TM}^{(0,-1)}(0)
e^{-2ak_{\bot}\sqrt{\varepsilon_{xx,0}^{(0)}}/\sqrt{\varepsilon_{zz,0}^{(0)}}}
\right],
\nonumber\\[-2mm]
&&\label{eq27}
\\
&&
{P}_{\rm cl}(a,T)=-\frac{k_BT}{2\pi}
\sqrt{\frac{\varepsilon_{xx,0}^{(0)}}{\varepsilon_{zz,0}^{(0)}}}
\int_{0}^{\infty} k_{\bot}^2dk_{\bot}
\left[\frac{e^{2ak_{\bot}
\sqrt{\varepsilon_{xx,0}^{(0)}}/\sqrt{\varepsilon_{zz,0}^{(0)}}}}{r_{\rm TM}^{(0,+1)}(0)
r_{\rm TM}^{(0,-1)}(0)}-1\right]^{-1}.
\nonumber
\end{eqnarray}
\noindent
Now it is convenient to introduce the new variable (\ref{eq17}) with $l=0$ and repeat
the same calculation as in Eqs.~(\ref{eq18})--(\ref{eq20}) with the result
\begin{eqnarray}
&&
{\cal F}_{\rm cl}(a,T)=-\frac{k_BT}{16\pi a^2}
\frac{\varepsilon_{zz,0}^{(0)}}{\varepsilon_{xx,0}^{(0)}}
{\rm Li}_{3}\left[r_{\rm TM}^{(0,+1)}(0)r_{\rm TM}^{(0,-1)}(0)\right],
\nonumber\\[-2mm]
&&\label{eq28}
\\
&&
{P}_{\rm cl}(a,T)=-\frac{k_BT}{8\pi a^3}
\frac{\varepsilon_{zz,0}^{(0)}}{\varepsilon_{xx,0}^{(0)}}
{\rm Li}_{3}\left[r_{\rm TM}^{(0,+1)}(0)r_{\rm TM}^{(0,-1)}(0)\right].
\nonumber
\end{eqnarray}
\noindent
As compared to the case of an isotropic material in the gap between plates,
here we have the additional factor
${\varepsilon_{zz,0}^{(0)}}/{\varepsilon_{xx,0}^{(0)}}$.

Similar situation holds when one plate is dielectric and another plate
is metallic. In this case the TE mode again does not contribute due to
$r_{\rm TE}^{(0,-1)}(0)=0$. The TM reflection coefficient for the dielectric
film, $r_{\rm TM}^{(0,-1)}(0)$, is again given by Eq.~(\ref{eq26}).
For metallic plate, one has from Eq.~(\ref{eq26})
$r_{\rm TM}^{(0,+1)}(0)=1$ because $\varepsilon_{0}^{(+1)}=\infty$.
As a result, the classical limit for two dissimilar plates, one dielectric
and another one metallic, with a uniaxial anisotropic film between them
takes the form
\begin{eqnarray}
&&
{\cal F}_{\rm cl}(a,T)=-\frac{k_BT}{16\pi a^2}
\frac{\varepsilon_{zz,0}^{(0)}}{\varepsilon_{xx,0}^{(0)}}
{\rm Li}_{3}\left[r_{\rm TM}^{(0,-1)}(0)\right],
\nonumber\\[-2mm]
&&\label{eq29}
 \\
&&
{P}_{\rm cl}(a,T)=-\frac{k_BT}{8\pi a^3}
\frac{\varepsilon_{zz,0}^{(0)}}{\varepsilon_{xx,0}^{(0)}}
{\rm Li}_{3}\left[r_{\rm TM}^{(0,-1)}(0)\right].
\nonumber
\end{eqnarray}

Finally we consider  the classical limit for two metallic plates with an
anisotropic film between them. In this case from Eq.~(\ref{eq3}) one has
$r_{\rm TM}^{(0,\pm 1)}(0)=1$.
As to the contribution of the TE mode, it is the same as for two metals
separated by the vacuum gap \cite{5} because the quantity
$k_{\rm TE}^{(0)}(0,k_{\bot})$ in the second line of Eq.~(\ref{eq2}) does
not depend on $\varepsilon_{xx,0}^{(0)}$.
Then, from Eq.~(\ref{eq1}) with account of Eq.~(\ref{eq2}) one obtains
\begin{equation}
{\cal F}_{\rm cl}(a,T)=-\frac{k_BT\zeta(3)}{16\pi a^2}\left[
\frac{\varepsilon_{zz,0}^{(0)}}{\varepsilon_{xx,0}^{(0)}}+1
-4\frac{\delta_0}{a}+12\left(\frac{\delta_0}{a}\right)^2\right],
\label{eq30}
\end{equation}
\noindent
where $\zeta(z)$ is the Riemann zeta function and
$\delta_0=\lambda_p/(2\pi)=c/\omega_p$ is the effective penetration depth
of the electromagnetic oscillations into a metal.
In a similar way, for the classical Casimir pressure we have
\begin{equation}
{P}_{\rm cl}(a,T)=-\frac{k_BT\zeta(3)}{8\pi a^3}\left[
\frac{\varepsilon_{zz,0}^{(0)}}{\varepsilon_{xx,0}^{(0)}}+1
-6\frac{\delta_0}{a}+24\left(\frac{\delta_0}{a}\right)^2\right].
\label{eq31}
\end{equation}

As can be seen from Eqs.~(\ref{eq30}) and (\ref{eq31}), there is
important qualitative difference between the classical Casimir free energy
and pressure in the configuration of two metallic plates separated by the
isotropic and anisotropic films. In the case of an isotropic film, the classical
Casimir effect does not depend on the material properties of the film.
It is the common property for all dielectric films independently of the values of their
static dielectric permittivity. By contrast, for uniaxial anisotropic films
the classical limit depends on the film material properties in accordance
to Eqs.~(\ref{eq30}) and (\ref{eq31}), i.e., through the value of the
ratio ${\varepsilon_{zz,0}^{(0)}}/{\varepsilon_{xx,0}^{(0)}}$.

In the end of this section we consider the measure of agreement between the
classical Casimir pressures in Eqs.~(\ref{eq28}), (\ref{eq29}) and (\ref{eq31})
and the exact values of the Casimir pressure computed numerically in Sec.~III.
The classical Casimir pressures calculated by the second lines
Eqs.~(\ref{eq28}), (\ref{eq29}), and by Eq.~(\ref{eq31}) are shown by the
dashed lines in Figs.~\ref{fg2}(a), \ref{fg3}, and \ref{fg2}(b), respectively.
The comparison with the respective solid lines shows that for two SiO${}_2$
plates the relative deviation between the magnitudes of the classical and
exact Casimir pressures
\begin{equation}
\delta P_{\rm cl}=\frac{|P_{\rm cl}|-|P|}{|P|}
\label{eq32}
\end{equation}
\noindent
is equal to --6\%, --1.97\%, --0.56\% and --0.15\% for film thicknesses
equal to 1.5, 2, 2.5, and $3\,\mu$m, respectively.
One can conclude that in this case the classical limit is achieved
for $a\approx 2\,\mu$m. In a like manner, for two dissimilar Au-SiO${}_2$
plates (Fig.~\ref{fg3}) $\delta P_{\rm cl}=5.4$\%, 1.4\%, 0.33\%, and 0.08\%
at $a=1.5$, 2, 2.5, and $3\,\mu$m, respectively.
{}From Fig.~\ref{fg2}(b) (Au-Au plates) one obtains
$\delta P_{\rm cl}=-18$\%, --6.4\%, --1.9\%, and --0.55\%
at the same respective film thicknesses (plate separations). Thus, for
Au-SiO${}_2$ and for Au-Au plates interacting across a uniaxial anisotropic
film the classical limit is achieved for film thicknesses equal to
approximately 2 and $2.5\,\mu$m, respectively.

\section{Conclusions and discussion}

In this paper, we have investigated the thermal Casimir force between two
dielectric, metallic and dissimilar (one dielectric and one metallic) plates
made of isotropic materials across a dielectric film made of a uniaxial
anisotropic crystal. Although the Lifshitz formula for the Casimir free
energy and pressure has been generalized for this case in the previous
literature, the most of calculations were performed in the nonrelativistic
limit, and specific features of the thermal Casimir force with account of
retardation effects and in the classical limit were not considered.

According to our results, in this case the role of relativistic effects
is much larger than for a vacuum gap, so that the standard theory of the
van der Waals force is not applicable even at the shortest separations
between the plates. We have obtained simple analytic expressions for the
Casimir free energy and pressure in the classical limit, i.e., at large
separations (high temperatures). In the case of two metallic plates
separated by an isotropic film the classical Casimir free energy and
pressure are known to be independent on the film material properties.
We demonstrated that this result is not preserved for anisotropic films,
where both the Casimir free energy and pressure depend on the ratio of
static dielectric permittivities of the film along different coordinate
axes. We have also compared the results of numerical computations using
the exact Lifshitz formula with the analytical results in the classical
limit. It was shown that for two isotropic plates separated by an
anisotropic film the classical limit is achieved at much shorter
separations than for the plates separated by a vacuum gap.

As was noted in Sec.~I, the configuration of a  uniaxial film
sandwiched between two isotropic plates is of much importance in the
investigation of stability of strongly confined (anisotropic) liquid
crystals. In this case one can replace the plates with a vacuum and
arrive to the free  energy of an anisotropic film alone.
Very recently, the Lifshitz formula adapted to the case of
two-dimensional structures found topical applications in the investigation
of the Casimir effect for graphene \cite{38,39,40,41,42}.
The most fundamental formalism describing interaction of graphene with the
electromagnetic fluctuations is based on the use of the polarization
tensor in (2+1)-dimensional space-time \cite{43,44,45}. In its turn,
this tensor is equivalent \cite{46} to two (nonlocal) dielectric
permittivities, the longitudinal one and the transverse one, in some
analogy to the uniaxial crystals considered in this paper.

One can conclude that the Casimir effect for layered structures, where some
of the layers are made of anisotropic materials, possesses some unusual properties
which can be potentially interesting for both fundamental physics and
nanotechnological applications.

\section*{Acknowledgments}

The author is grateful to G.~L.~Klimchitskaya for helpful
discussions.


\begin{figure}[b]
\vspace*{-8cm}
\centerline{\hspace*{1cm}
\includegraphics{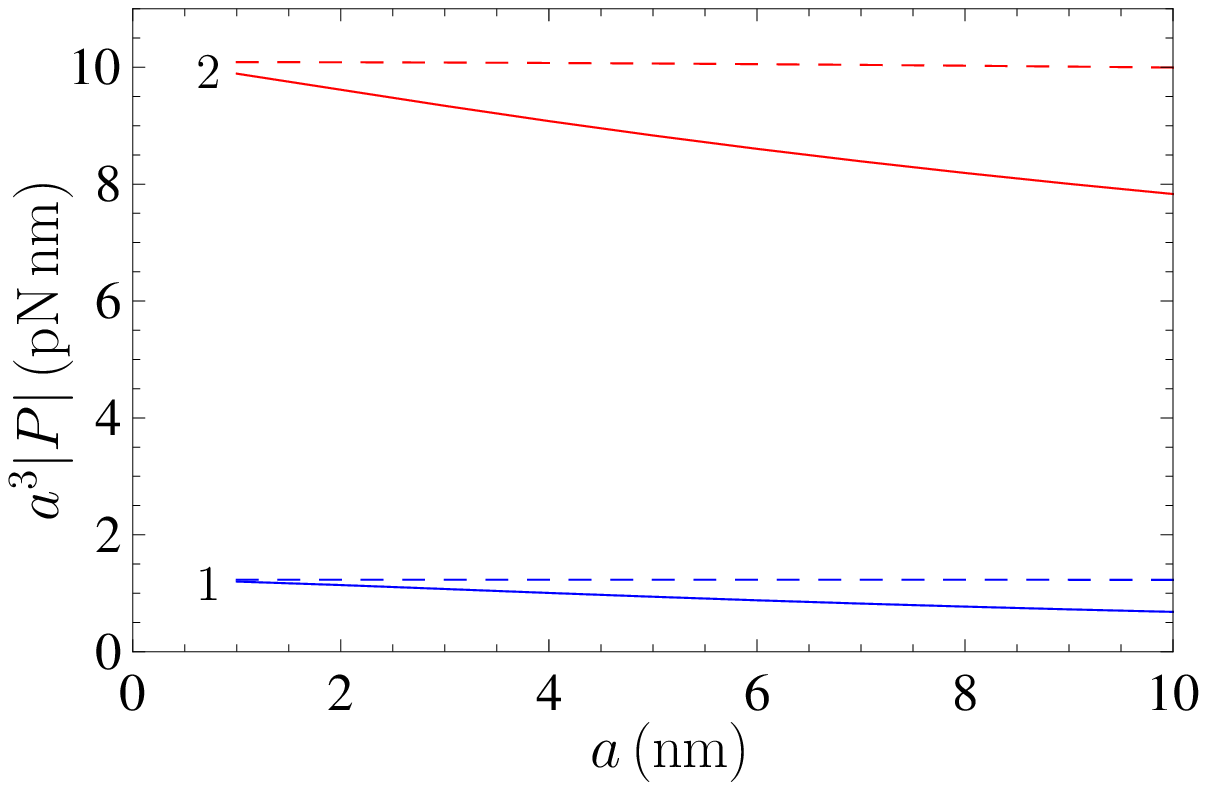}
}
\vspace*{-9cm}
\caption{\label{fg1}(Color online)
The  magnitudes of the Casimir pressure between two
parallel isotropic plates
 interacting through a uniaxial anisotropic film (BeO)
multiplied by $a^3$ are
 calculated at $T=300\,$K as functions of film thickness
using the exact Lifshitz formula (the solid lines)
and in the nonrelativistic limit (the dashed lines).
The lines 1 and 2 are plotted for the case of SiO${}_2$
and Au plates, respectively.
}
\end{figure}
\begin{figure}[b]
\vspace*{-3cm}
\centerline{\hspace*{1cm}
\includegraphics{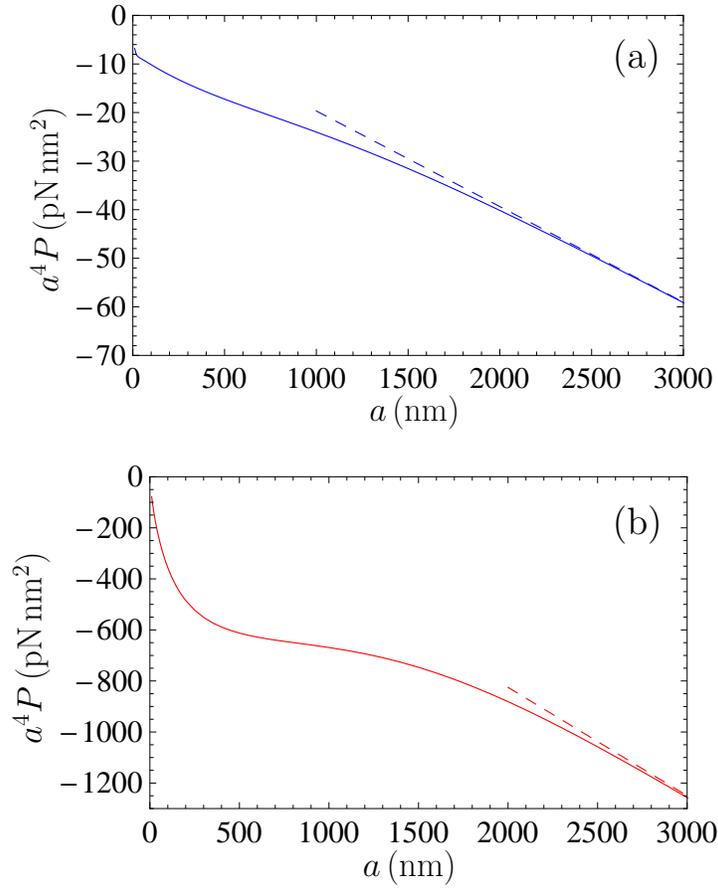}
}
\vspace*{-13cm}
\caption{\label{fg2}(Color online)
The  Casimir pressure between two parallel isotropic plates
 interacting through a uniaxial anisotropic film (BeO)
multiplied by $a^4$ are
 calculated at $T=300\,$K as functions of film thickness
(the solid lines). The dashed lines show the classical limit.
(a) The plates are made of SiO${}_2$.
(b) The plates are made of Au.
}
\end{figure}
\begin{figure}[b]
\vspace*{-8cm}
\centerline{\hspace*{1cm}
\includegraphics{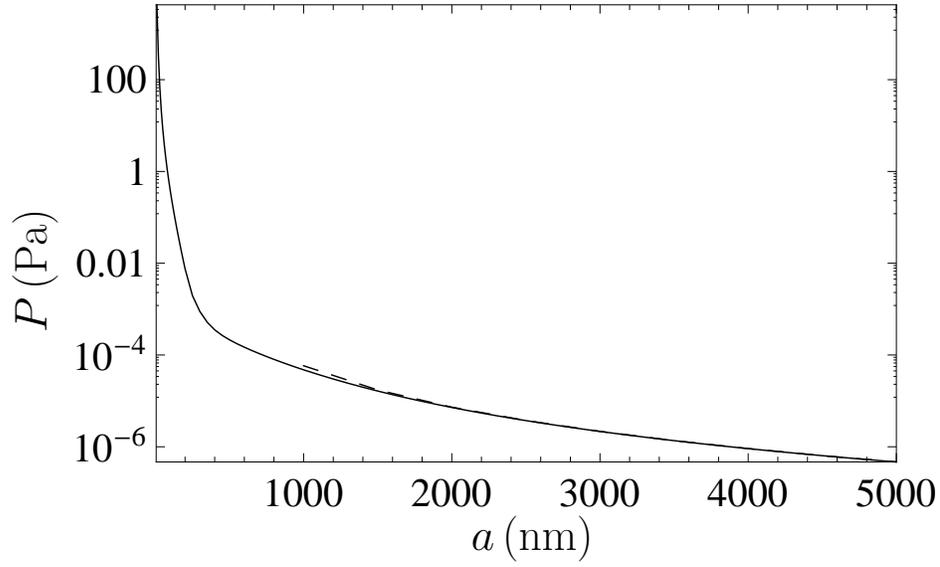}
}
\vspace*{-9cm}
\caption{\label{fg3}
The  Casimir pressure between two dissimilar isotropic plates
made of SiO${}_2$ and Au interacting through a uniaxial anisotropic
film (BeO) is calculated at $T=300\,$K as a function of film thickness
(the solid line). The dashed line shows the classical limit.
}
\end{figure}
\end{document}